\documentclass[preprint,aps,12pt,showpacs,nofootinbib,tightenlines]{revtex4}
\usepackage{amsmath}
\usepackage{amssymb}
\usepackage{epsfig}
\usepackage{graphicx}
\begin{document}

\newcommand{\beq}{\begin{eqnarray}}
\newcommand{\eeq}{\end{eqnarray}}
\newcommand{\non}{\nonumber\\ }

\newcommand{\acp}{{\cal A}_{CP}}
\newcommand{\calm}{{\cal M} }
\newcommand{\etap}{\eta^{(\prime)} }
\newcommand{\etapr}{\eta^\prime }
\newcommand{\jpsi}{ J/\Psi }
\newcommand{\calh}{ {\cal H} }

\newcommand{\pb}{\phi_{B_s}}
\newcommand{\pp}{\phi_{\pi}}
\newcommand{\pe}{\phi_{\eta}^A}
\newcommand{\pepr}{\phi_{\eta^\prime}^A}
\newcommand{\ppp}{\phi_{\pi}^P}
\newcommand{\pep}{\phi_{\eta}^P}
\newcommand{\peprp}{\phi_{\eta^\prime}^P}
\newcommand{\ppt}{\phi_{\pi}^t}
\newcommand{\pet}{\phi_{\eta}^T}
\newcommand{\peprt}{\phi_{\eta^\prime}^T}
\newcommand{\fb}{f_{B_s} }
\newcommand{\fpi}{f_{\pi} }
\newcommand{\feta}{f_{\eta} }
\newcommand{\fetap}{f_{\eta^\prime} }
\newcommand{\rpi}{r_{\pi} }
\newcommand{\re}{r_{\eta} }
\newcommand{\rep}{r_{\eta^\prime} }
\newcommand{\mbs}{m_{B_s} }
\newcommand{\mw}{m_{B_s} }
\newcommand{\mop}{m_{0\pi} }
\newcommand{\moe}{m_{0\eta} }
\newcommand{\moep}{m_{0\eta^\prime} }

\newcommand{\psl}{ P \hspace{-2.8truemm}/ }
\newcommand{\nsl}{ n \hspace{-2.2truemm}/ }
\newcommand{\vsl}{ v \hspace{-2.2truemm}/ }
\newcommand{\epsl}{\epsilon \hspace{-1.8truemm}/\,  }
\def \ctp{ Commun.Theor.Phys. }
\def \epjc{ Eur.Phys.J. C }
\def \ijmpa{ Int.J.Mod.Phys. A }
\def \jpg{  J. Phys. G }
\def \npb{  Nucl. Phys. B }
\def \plb{  Phys. Lett. B }
\def \pr{  Phys. Rep. }
\def \prd{  Phys. Rev. D }
\def \prl{  Phys. Rev. Lett.  }
\def \zpc{  Z. Phys. C  }
\def \jhep{ J. High Energy Phys.  }

\title{$B \to (\jpsi,\eta_c) K$ decays in the perturbative QCD approach}
\author{Xin Liu\footnote{ hsinlau@126.com},
Zhi-Qing Zhang \footnote{zhangzhiqing@zzu.edu.cn}
and Zhen-Jun Xiao \footnote{ xiaozhenjun@njnu.edu.cn }}
\affiliation{ Department of Physics and Institute of Theoretical Physics,
Nanjing Normal University, Nanjing, Jiangsu 210097, P.R. China} 

\date{\today}
\begin{abstract}
In this paper, we calculated the $B \to (\jpsi, \eta_c) K$ decays in
the perturbative QCD approach with the inclusion of the partial
next-to-leading order (NLO) contributions. We found  that
(a) when the large enhancements from the known NLO contributions are taken into account, the NLO pQCD predictions 
for the branching ratios are the following: 
$Br(B^0 \to \jpsi K^0) = 5.2^{+3.5}_{-2.8}\times 10^{-4}$,
$Br(B^+ \to \jpsi K^+) = 5.6^{+3.7}_{-2.9}\times 10^{-4}$, 
$Br(B^0 \to \eta_c K^0) = 5.5^{+2.3}_{-2.0}\times 10^{-4}$,
$Br(B^+ \to \eta_c K^+) = 5.9^{+2.5}_{-2.1}\times 10^{-4}$,  
which are roughly $40\%$ smaller than the measured values, but basically agree with the data 
within $2-\sigma$ errors;  
(b) the NLO pQCD predictions for the CP-violating asymmetries of $B
\to (\jpsi,\eta_c)K$ decays agree perfectly with the data.
\end{abstract}

\pacs{13.25.Hw, 12.38.Bx, 14.40.Nd}

\maketitle

\section{Introduction}\label{sec:s1}

The $B \to \jpsi K$ and $B \to \eta_c K$ decays are phenomenologically very interesting
decay modes and have drawn a great attention for many years.
Although the underlaying weak decay of $b \to c \bar{c} s$ is simple,
but a clear understanding of  the exclusive $B \to (\jpsi, \eta_c) K$ decays
is really difficult because of the involving of the complex strong-interaction
effects.

On the experiments side, the experimental studies for the
"Golden-plated" $B\to \jpsi K_{S,L}$ decays result in the precision measurement
of $\sin{2\beta}$\cite{pdg2008}. The branching ratios of $B\to (\jpsi,\eta_c) K$ decays
and other similar decays involving a charmonium and a light pseudo-scalar or vector meson
as the two final state meson, have been measured with good or high precision
\cite{pdg2008,hfag2008}:
\beq
Br(B^0\to \jpsi K^0) &=& (8.71 \pm 0.32) \times 10^{-4}, \non
Br(B^+\to \jpsi K^+) &=& (10.07 \pm 0.35) \times 10^{-4}, \label{eq:exp1}\\
Br(B^0\to \eta_c K^0) &=& (8.9 \pm 1.6) \times 10^{-4}, \non
Br(B^+\to \eta_c K^+) &=& (9.1 \pm 1.3) \times 10^{-4}.
\label{eq:exp2}
\eeq
The accuracy of above measurements will be improved rapidly
along with the running of the relevant LHC experiments.

On the theory side,  such B meson charmonia decays have been studied
intensively by employing various theoretical methods or approaches, for example, in
Refs.~\cite{gkp,cheng99,ck00,cheng01,boos04,chao1,chao2,melic,cl97,chen05,li07a}.
But unfortunately, it is still very difficult to give an
satisfactory explanation for the corresponding data without the worry about
the serious problems.

For $B\to \jpsi K$ decay, for example, the theoretical predictions
for its branching ratio in both the naive factorization approach
(NFA) \cite{cheng99} and the QCD factorization (QCDF)
approach\cite{bbns99} are much smaller (a factor of $7\sim 10$ )
than the measured values\cite{ck00}: $Br(B \to \jpsi K) \sim
1.1\times 10^{-4}$ when the twist-2 distribution function (DA)
$\phi_{\jpsi}(x) = 6x (1-x)$ was employed \cite{ck00}.

In Ref.~\cite{cheng01}, the authors studied the effects of twist-3
DA $\phi_\sigma^K$ and found that the resultant enhancement to the
Wilson coefficient $a_2(\jpsi K)$ and consequently to the branching
ratio $Br(B \to \jpsi K)$, induced through the spectator diagram,
can be large. But one should note that there are also logarithmic
divergences arising from spectator interactions due to kaon twist-3
effects, this is always a serious problem in the QCDF approach.

In Refs.~\cite{chao1,chao2}, the authors studied the decays $B\to
(\eta_c, \eta_c^\prime, \chi_{c0}, \chi_{c1}) K$ in the QCDF approach
and found that (a) the logarithmic divergences will arise from the
spectator interactions due to the kaon twist-3 effects; and (b) the
predicted decay rate is $Br(B \to \eta_c K) =1.9\times 10^{-4}$,
which is still  a factor of 5 smaller than the measured value
in Eq.(\ref{eq:exp2}). They concluded that the QCDF
approach with its present version can not be safely applied to
exclusive decays of B meson into charmonia \cite{chao2}.

The $B \to \jpsi K$ decays have also been investigated by employing
the QCD light-cone sum rules (LCSR) \cite{melic}. The authors
calculated the nonfactorizable contributions to the $B \to \jpsi K$
decay coming from the exchanges of the soft gluons between the
emitted $\jpsi$ and the kaon. But their predictions for branching
ratios is still too small, $Br(B \to \jpsi K) \sim 3.3  \times
10^{-4}$, to accommodate the data.

In Refs.\cite{chen05,li07a}, the authors studied $B \to (\jpsi, \eta_c, \chi_{c0,c1}) K^{(*)}$
decays in a formalism that combines the QCDF factorization and the perturbtive QCD (pQCD)
approaches\cite{cl97}. They employed the QCDF approach to calculate the factorizable
contribution, but the pQCD approach to evaluate the nonfactorizable corrections to the
considered decays. According to their studies\cite{chen05,li07a} we see that (a) the
theoretical predictions for the branching ratios of $B \to (\jpsi, \chi_{c0,c1}) K$ decays
can be large and consistent with the data; (b) the $B \to \eta_c K$ decays still exhibit
a puzzle: the predicted result is $Br(B \to \eta_c K) \approx 2.3\times 10^{-4}$, much
smaller than the measured values as given in Eq.~(\ref{eq:exp1}).
Furthermore, it should be mentioned that these
results\cite{chen05,li07a} were obtained by treating one decay with two different
factorization approaches: the self-consistency of such ``mixing-approach"
may be  a serious problem.

Up to now, a clear and satisfactory theoretical interpretation for
the measured large decay rates of $B \to (\jpsi, \eta_c ) K$ are
still absent. We call this situation the ``$B\to (\jpsi, \eta_c) K$"
puzzle. In this paper, we will calculate the branching ratios and
CP-violating asymmetries of the four $B \to (J/\Psi,\eta_c) K$
decays by employing the pQCD factorization approach: (a) we evaluate
both the factorizable and nonfactorizable contributions in the pQCD
approach; (b) besides the full leading order (LO) contributions in
the pQCD approach, the currently known next-to-leading order (NLO)
contributions \cite{nlo05} (specifically the QCD vertex corrections
for the considered decays) are also included.

The paper is organized as follows: in Sec.~\ref{sec:s2},
we firstly present the formalism of the pQCD approach, and then
make the analytic calculations and show the decay amplitudes for the considered
decays. In Sec.~\ref{sec:s3}, we show the numerical results and compare
them with the measured values. A short summary and some conclusions are given
in the last section.

\section{Formalism and Perturbative Calculations}\label{sec:s2}

\subsection{Formalism}

In recent years, the pQCD factorization approach has been used
frequently to calculate various B meson decay channels. For the two
body charmless hadronic B meson decays the pQCD predictions for the
branching ratios and CP-violating asymmetries generally agree well
with the measured values
\cite{li2001,li2003,xiao06,ali07,xiao08,xiao08b}. In
Ref.~\cite{ll03}, the authors calculated $B \to D_s^* K, D_s^{(*)+}
D_s^{(*)-}$ and $B_s \to  D^{(*)+} D^{(*)-}$ decays and found that
the pQCD approach works well for such decays. In a previous
paper\cite{jpsi08}, the $B \to \jpsi K$ decays have been studied by
employing the pQCD approach at leading order. Here we try to apply
the pQCD approach to calculate the $B\to (\jpsi, \eta_c) K $ decays
with the inclusion of the NLO corrections.

In pQCD approach, the decay amplitude of $B \to M_2 M_3$
decays\footnote{Here $M_2=(\jpsi,\eta_c)$ is the emitted charmonium,
and $M_3$ is the kaon which absorbed the spectator quark. }
can be written conceptually as the convolution,
\beq
{\cal A}(B \to M_2 M_3)\sim \int\!\! d^4k_1 d^4k_2 d^4k_3\ \mathrm{Tr} \left [ C(t)
\Phi_B(k_1) \Phi_{M_2}(k_2) \Phi_{M_3}(k_3) H(k_1,k_2,k_3, t) \right ],
\label{eq:con1}
\eeq
where the term ``$\mathrm{Tr}$" denotes the trace over Dirac and color indices.
$C(t)$ is the Wilson
coefficient which results from the radiative corrections at short
distance. In the above convolution, $C(t)$ includes the harder
dynamics at larger scale than $m_B$ scale and describes the
evolution of local $4$-Fermi operators from $m_W$ (the $W$ boson
mass) down to $t\sim\mathcal{O}(\sqrt{\bar{\Lambda} m_B} )$
scale, where $\bar{\Lambda}\equiv m_B -m_b$. The function
$H(k_1,k_2,k_3,t)$ is the hard part and can be calculated
perturbatively. The function $\Phi_M$ is the wave function which
describes hadronization of the quark and anti-quark to the meson
$M$. While the function $H$ depends on the process considered, the
wave function $\Phi_M$ is independent of the specific process. Using
the wave functions determined from other well measured processes,
one can make quantitative predictions here.

Using the light-cone coordinates the $B$ meson and the two final
state meson momenta can be written as
\beq P_1 =\frac{m_B}{\sqrt{2}}
(1,1,{\bf 0}_T), \quad P_2 =\frac{m_B}{\sqrt{2}} (1,r^2,{\bf 0}_T),
\quad P_3 =\frac{m_B}{\sqrt{2}} (0,1-r^2,{\bf 0}_T),
\eeq
respectively, where $r=m_{M_2}/m_B$, and the light pseudo-scalar
meson masses $m_{M_3}=m_K$ have been neglected. The longitudinal
polarization of vector $J/\Psi$, $\epsilon_L$, is given by
$\epsilon_L = \frac{m_B}{\sqrt{2}m_{\jpsi}} (1, -r_{\jpsi}^2,{\bf
0}_T)$. Putting the light (anti-) quark momenta in $B$ and $M_3$
mesons as $k_1$ and $k_3$, respectively, we can choose
\beq
k_1 = (x_1 P_1^+,0,{\bf k}_{1T}),  \quad
k_3 = (0, x_3 P_3^-,{\bf k}_{3T}).
\eeq
For $M_2$, the momentum fraction of c quark  is
chosen as $x_2 P_2$. Then, the integration over $k_1$, $k_2$, and
$k_3$ in Eq.(\ref{eq:con1}) will lead to
\beq
{\cal A}(B \to M_2
M_3) &\sim &\int\!\! d x_1 d x_2 d x_3 b_1 d b_1 b_2 d b_2 b_3 d b_3
\non && \cdot \mathrm{Tr} \left [ C(t) \Phi_B(x_1,b_1)
\Phi_{M_2}(x_2,b_2) \Phi_{M_3}(x_3, b_3) H(x_i, b_i, t) S_t(x_i)\,
e^{-S(t)} \right ], \ \  \label{eq:am2m3}
\eeq
where $b_i$ is the
conjugate space coordinate of $k_{iT}$, and $t$ is the largest
energy scale in function $H(x_i,b_i,t)$. The large logarithms $\ln
(m_W/t)$ are included in the Wilson coefficients $C(t)$. The large
double logarithms ($\ln^2 x_i$) on the longitudinal direction are
summed by the threshold resummation ~\cite{li02}, and they lead to
$S_t(x_i)$ which smears the end-point singularities on $x_i$. The
last term, $e^{-S(t)}$, is the Sudakov form factor which suppresses
the soft dynamics effectively ~\cite{soft}. Thus it makes the
perturbative calculation of the hard part $H$ applicable at
intermediate scale, i.e., $m_B$ scale.

For the considered decays, the weak effective Hamiltonian
$\calh_{eff}$ for $b \to s$ transition can be written as
\beq
\label{eq:heff} {\cal H}_{eff} = \frac{G_{F}} {\sqrt{2}} \,
\left[ V_{cb}^* V_{cs} \left (C_1(\mu) O_1^c(\mu) + C_2(\mu) O_2^c(\mu)
\right) - V_{tb}^* V_{ts} \, \sum_{i=3}^{10} C_{i}(\mu) \,O_i(\mu)
\right] \; ,
\eeq
where $C_i(\mu)$ are Wilson coefficients at the
renormalization scale $\mu$ and $O_i$ are the four-fermion
operators:
\beq
\begin{array}{llllll}
O_1^{c} & = &   \bar s_\alpha\gamma^\mu L c_\beta\cdot \bar
c_\beta\gamma_\mu L b_\alpha\ , & O_2^{c} & = &\bar
s_\alpha\gamma^\mu L c_\alpha\cdot \bar
c_\beta\gamma_\mu L b_\beta\ , \\
O_3 & = & \bar s_\alpha\gamma^\mu L b_\alpha\cdot \sum_{q'}\bar
 q_\beta'\gamma_\mu L q_\beta'\ ,   &
O_4 & = & \bar s_\alpha \gamma^\mu L b_\beta\cdot
\sum_{q'}\bar q_\beta'\gamma_\mu L q_\alpha'\ , \\
O_5 & = & \bar s_\alpha\gamma^\mu L b_\alpha\cdot \sum_{q'}\bar
q_\beta'\gamma_\mu R q_\beta'\ ,   & O_6 & = & \bar
s_\alpha\gamma^\mu L b_\beta\cdot \sum_{q'}\bar q_\beta'\gamma_\mu R q_\alpha'\ , \\
O_7 & = & \frac{3}{2}\bar s_\alpha \gamma^\mu L b_\alpha\cdot
\sum_{q'}e_{q'}\bar q_\beta'\gamma_\mu R q_\beta'\ , & O_8 & = &
\frac{3}{2}\bar s_\alpha \gamma^\mu L b_\beta\cdot
\sum_{q'}e_{q'}\bar q_\beta'\gamma_\mu R q_\alpha'\ , \\
O_9 & = & \frac{3}{2}\bar s_\alpha \gamma^\mu L b_\alpha\cdot
\sum_{q'}e_{q'}\bar q_\beta'\gamma_\mu L q_\beta'\ , & O_{10} & = &
\frac{3}{2}\bar s_\alpha \gamma^\mu L b_\beta\cdot
\sum_{q'}e_{q'}\bar q_\beta'\gamma_\mu L q_\alpha'\ ,
\label{eq:operators}
\end{array}
\eeq where $\alpha$ and $\beta$ are the $SU(3)$ color indices; $L$
and $R$ are the left- and right-handed projection operators with
$L=(1 - \gamma_5)$, $R= (1 + \gamma_5)$. The sum over $q'$ runs over
the quark fields that are active at the scale $\mu=O(m_b)$, i.e.,
$q'\epsilon\{u,d,s,c,b\}$.

In PQCD approach, the scale $``t"$ appeared in the Wilson
coefficients $C_i(t)$, the hard-kernel $H(x_i,b_i,t)$ and the
Sudakov factor $e^{-S(t)}$ is chosen as the largest energy scale in
the gluon and/or the quark propagators of a given Feynman diagram,
in order to suppress the higher order corrections and improve the
reliability of the perturbative calculation. Here, the scale $``t"$
may be larger or smaller than the $m_b$ scale. In the range of $ t <
m_b $ or $t \geq m_b$, the number of active quarks is $N_f=4$ or
$N_f=5$, respectively. For the Wilson coefficients $C_i(\mu)$ and
their renormalization group (RG) running, they are known at NLO
level currently \cite{buras96}. The explicit expressions of the LO
and NLO $C_i(m_B)$ can be found easily, for example, in
Refs.~\cite{buras96,luy01}.

When the pQCD approach at leading-order are employed, the leading order Wilson
coefficients $C_i(m_W)$, the leading order RG evolution matrix $U(t,m)^{(0)}$ from
the high scale $m$ down to $t < m$ and the leading order $\alpha_s(t)$ are used:
\beq
\alpha_s(t)=\frac{4\pi}{ \beta_0 \ln \left [ t^2/ \Lambda_{QCD}^2\right]},
\eeq
where $\beta_0 = (33- 2 N_f)/3$.

When the NLO contributions are taken into account, however,
the NLO Wilson coefficients $C_i(m_W)$, the NLO RG evolution matrix $U(t,m,\alpha)$
( see Eq.~(7.22) in Ref.~\cite{buras96}) and the $\alpha_s(t)$ at two-loop level
will be used:
\beq
\alpha_s(t)=\frac{4\pi}{ \beta_0 \ln \left [ t^2/ \Lambda_{QCD}^2\right]}
\cdot \left \{ 1- \frac{\beta_1}{\beta_0^2 } \cdot
\frac{ \ln\left [ \ln\left [ t^2/\Lambda_{QCD}^2  \right]\right]}{
\ln\left [ t^2/\Lambda_{QCD}^2\right]} \right \},
\label{eq:asnlo}
\eeq
where $\beta_0 = (33- 2 N_f)/3$, $\beta_1 = (306-38 N_f)/3$. By assuming
$\Lambda_{QCD}^{(5)}=0.225$ GeV, we will get $\Lambda_{QCD}^{(4)}=0.287$ GeV ($0.326$ GeV)
for LO (NLO) case.

As discussed in Ref.\cite{xiao08}, it is reasonable to choose $\mu_0=1.0$ GeV
as the lower cut-off of the hard scale $t$.
In the numerical integrations we will
fix the values $C_{i}(t)$ at $C_{i}(1.0)$  whenever the scale $t$ runs below the
scale $\mu_0=1.0$ GeV \cite{xiao08,xiao08b}, unless otherwise stated.

\begin{figure}[tb]
\vspace{-6cm} \centerline{\epsfxsize=20 cm \epsffile{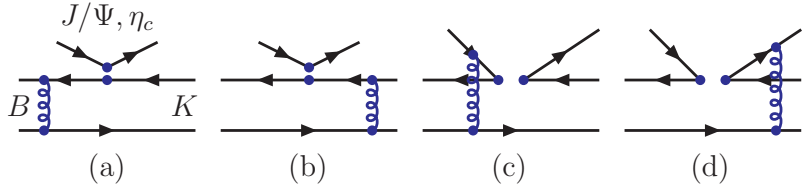}}
\vspace{-20cm} \caption{Typical Feynman diagrams contributing to  $B
\to (J/\Psi,\eta_c) K$ decays at leading order.}
 \label{fig:fig1}
\end{figure}

\subsection{ $B \to J/\Psi M_3 $ decays at leading order}\label{ssec:jm3}

At the leading order pQCD approach, as illustrated in Fig.~1, the relevant Feynman diagrams
for the considered decays include the factorizable emission diagrams (Figs.1a and 1b)
and the non-factorizable spectator ones (Figs.1c and 1d).
The operators $O_{1,2}$, $O_{3,4}$ and $O_{9,10}$ are the $(V-A)(V-A)$ currents, while
$O_{5,6}$ and $O_{7,8}$ are the $(V-A)(V+A)$ currents. By
analytic calculations of Fig.1a and 1b, one finds the corresponding
decay amplitudes
\beq
F_{J/\Psi M_3}^{V-A}&=& 8 \pi C_F m_B^2
\int_0^1 d x_{1} dx_{3}\, \int_{0}^{\infty} b_1 db_1 b_3 db_3\,
\phi_{B}(x_1,b_1) \non & & \times \biggl\{ \biggl[\left[( 1-r^2 )
(1+x_3) -x_3 r^2 \right] \phi_{M_3}^A(x_3) + r_0 (1-2x_3)\left[
\phi_{M_3}^P(x_3) +\phi_{M_3}^T(x_3) \right] \non && - r_0 r^2
\left[( 1 - 2 x_3 )\phi_{M_3}^P(x_3) - (1 + 2 x_3 )\phi_{M_3}^T(x_3)
\right]\biggr] \non && \cdot \alpha_s(t_e^1)\,
h_e(x_1,x_3,b_1,b_3)\exp\left [-S_{ab}(t_e^1)\right ] \non && + 2
r_0 \left(1- r^2\right) \phi_{M_3}^P (x_3) \cdot
\alpha_s(t_e^2)h_e(x_3,x_1,b_3,b_1)\exp\left[-S_{ab}(t_e^2)\right]
\biggr\},
\label{eq:ab}
\eeq
where $r_0=m_0^K/m_B$, and  $C_F=4/3$ is a color factor.
The hard function $h_e$, the scales $t_e^i$
and the Sudakov factors $S_{ab}$ are displayed in Appendix
\ref{sec:app1}.

Now we consider the contributions of the operators $O_{5,6,7,8}$ in
the Fig.1. In some
decay channels, some of these operators contribute to the decay
amplitude in a factorizable way. Since only the vector part of
$(V+A)$ current contribute to the vector meson production, $ \langle
M_3 |V-A|B\rangle \langle J/\Psi |V+A | 0 \rangle = \langle M_3 |V-A
|B \rangle \langle J/\Psi |V-A|0 \rangle,$ that is
 \beq
 F_{J/\Psi M_3}^{V+A}= F_{J/\Psi M_3}^{V-A} \; \label{eq:ab1} .
 \eeq

For the non-factorizable diagrams 1(c) and 1(d), all three meson
wave functions are involved. The integration of $b_3$ can be
performed using $\delta$ function $\delta(b_3-b_1)$, leaving only
integration of $b_1$ and $b_2$. For the $(V-A)(V-A)$ operators, the
corresponding decay amplitude is
\beq
 M_{J/\Psi M_3}^{V-A}&=& - \frac{16 \sqrt{6}}{3} \pi C_F m_B^2
\int_{0}^{1}d x_{1}d x_{2}\,d x_{3}\,\int_{0}^{\infty} b_1d b_1 b_2d
b_2\, \phi_{B}(x_1,b_1) \non
 & &\times
\biggl \{2 r r_c \phi_{J/\Psi}^t(x_2)\phi_{M_3}^A(x_3)-4 r r_0 r_c
\phi_{J/\Psi}^t(x_2)\; \phi_{M_3}^T(x_3)
   \non &&
 -\left[x_3 + 2 (x_2 - x_3 )r^2\right]\phi_{J/\Psi}^L(x_2)
\phi_{M_3}^A(x_3)  \non
 & &   +2 r_0 \left[x_3 + (2 x_2- x_3) r^2\right]\phi_{J/\Psi}^L(x_2)\phi_{M_3}^T(x_3)\biggr\}
 \non &&
 \cdot \alpha_s(t_f) h_f(x_1,x_2,x_3,b_1,b_2)\exp\left [-S_{cd}(t_f)\right ]
 \;,  \label{eq:cd}
\eeq
where $r_c=m_{c}/m_{B}$ and $m_c$ is the mass for $c$ quark.

For some decay channels, the $(S-P)(S+P)$ operators can be obtained from  the $(V-A)(V+A)$
operators  by making the Fierz transformation, in order to get right color and flavor
structure for factorization to work. For these $(S-P)(S+P)$ operators, the corresponding
decay amplitude can be written as
\beq
M_{J/\Psi M_3}^{S+P}&=& -M_{J/\Psi M_3}^{V-A}.
\label{eq:cd1}
\eeq


For $B \to J/\Psi  M_3$ decays, by combining the contributions from
different Feynman diagrams, the total decay amplitude can be written
as
\beq
{\cal M}(B \to J/\Psi M_3) &=& F_{J/\Psi M_3}^{V-A}
f_{J/\Psi} \left\{ V_{cb}^*V_{cs}\; a_2 -V_{tb}^{*}V_{ts} \left (
a_3+a_5+a_7 +a_9 \right) \right\} \non &&
 + M_{J/\Psi M_3}^{V-A} \left \{V_{cb}^*V_{cs} C_2
 - V_{tb}^*V_{ts}  \left (C_4-C_6-C_8+C_{10}\right)\right\},
 \label{eq:jpsim3}
\eeq
where $a_i$ is the combination of the Wilson coefficients
$C_i$:
\beq
a_2 &=& C_1 + \frac{C_2}{3}; \quad a_i=C_i +
\frac{C_{i+1}}{3}, \ \ for \ \ i=3,5,7,9,
\label{eq:ai}
\eeq
where $C_2 \sim 1$ is the largest one among all Wilson coefficients.

\subsection{ $B \to \eta_{c} M_3 $ decays at leading order}\label{ssec:etacm3}

Following the same procedure as for $B \to \jpsi K$ decays,
it is straightforward to calculate the decay amplitudes for
$B \to \eta_{c} M_3 $ decays.
\beq
{\cal M}(B \to \eta_c M_3) &=& F_{\eta_c M_3}^{V-A} f_{\eta_c} \left[ V_{cb}^*V_{cs}\; a_2
-V_{tb}^{*}V_{ts}\left ( a_3+a_5+a_7+a_9\right) \right ] \non  & & +
M_{\eta_c M_3}^{V-A} \left [ V_{cb}^*V_{cs} C_2 - V_{tb}^*V_{ts}
\left(C_4+C_6+C_8+C_{10}\right) \right], \label{eq:etacm3}
\eeq
where the functions $F_{\eta_c M_3}^{V-A}, M_{\eta_c M_3}^{V-A}$, etc,  are
of the form
\beq
F_{\eta_c M_3}^{V-A}&=& - F_{\eta_c M_3}^{V+A} = 8
 \pi C_F m_B^2 \int_0^1 d x_{1} dx_{3}\, \int_{0}^{\infty} b_1
db_1 b_3 db_3\, \phi_{B}(x_1,b_1) \non & & \times \biggl\{
\biggl[\left[( 1-r^2 ) (1+x_3) -x_3 r^2 \right] \phi_{M_3}^A(x_3) +
r_0 (1-2x_3)\left[ \phi_{M_3}^P(x_3) +\phi_{M_3}^T(x_3) \right] \non
&& + r_0 r^2 \left[(1 + 2 x_3 )\phi_{M_3}^P(x_3) - ( 1 - 2 x_3
)\phi_{M_3}^T(x_3) \right]\biggr] \non && \cdot \alpha_s(t_e^1)\,
h_e(x_1,x_3,b_1,b_3)\exp[-S_{ab}(t_e^1)] \non && + 2 r_0 \left(1-
r^2\right) \phi_{M_3}^P (x_3)\cdot
\alpha_s(t_e^2)h_e(x_3,x_1,b_3,b_1)\exp[-S_{ab}(t_e^2)]
\biggr\},
\eeq
\beq
M_{\eta_c M_3}^{V-A}&=&  M_{\eta_c M_3}^{S+P}=
-\frac{16 \sqrt{6}}{3} \pi C_F m_B^2 \int_{0}^{1}d x_{1}d x_{2}\,d
x_{3}\,\int_{0}^{\infty} b_1d b_1 b_2d b_2\, \phi_{B}(x_1,b_1)
\phi_{\eta_c}^v(x_2,b_2) \non
 & &\times x_3
\biggl [  \left (1-2 r^2 \right )\phi_{M_3}^A(x_3) - 2 r_0 \left (1-
r^2 \right )\phi_{M_3}^T(x_3)\biggr ] \non&& \cdot \alpha_s(t_f)
h_f(x_1,x_2,x_3,b_1,b_2)\exp[-S_{cd}(t_f)] \;,
\eeq
where $\phi_{\eta_c}^v$ is the leading twist-2 part of the distribution amplitude for the
pseudo-scalar meson $\eta_c$.

\subsection{ NLO contributions in pQCD approach}

\begin{figure}[tb]
\vspace{-6cm} \centerline{\epsfxsize=20 cm \epsffile{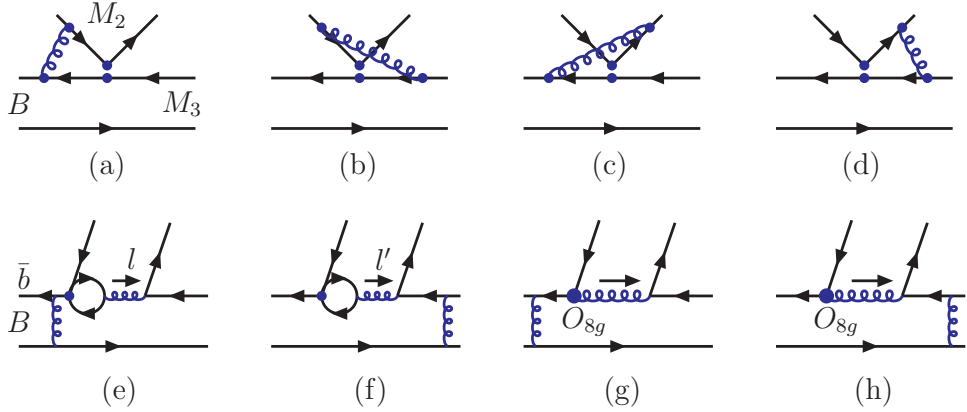}}
\vspace{-17cm} \caption{Typical Feynman diagrams contributing to  $B
\to M_2 M_3$ decays at NLO level.}
 \label{fig:fig2}
\end{figure}

For a general $B \to M_2 M_3$ decays, the power counting in the pQCD factorization
approach \cite{nlo05} is different from that in the QCD factorization\cite{bbns99}.
In the pQCD approach, the NLO contributions may include the following
parts\cite{xiao08,nlo05}:
\begin{enumerate}
\item
The Wilson coefficients $C_i(m_B)$ and the renormalization group
evolution matrix $U(t,m,\alpha)$ at the NLO level, and the $\alpha_s(t)$
at the two-loop level  \cite{buras96} should  be used.

\item
All the Feynman diagrams,  which lead to the decay amplitudes proportional to
$\alpha^2_s(t)$, should be considered.

\item
Currently known NLO contributions: (a) the vertex corrections;
(b) the contributions from the quark-loops and the chromo-magnetic penguins ($O_{8g}$),
as illustrated  in Fig.~\ref{fig:fig2}.

\item
The NLO contributions can also come from the Feynman diagrams as shown in
the Figs.~5-7 in Ref.~\cite{xiao08}.
The analytical calculations for these (more than 100!) Feynman diagrams have
not been completed yet.

\end{enumerate}

For the considered $B \to \jpsi K$ and $\eta_c K$ decays, only the
vertex corrections (see Fig.2a-2d) among the known NLO contributions
will contribute. For the four vertex correction diagrams Fig.2a-2d,
as was confirmed in Ref.\cite{cheng01}, the infrared divergences
from the soft gluons and collinear gluons in the four diagrams  will
be canceled each other, respectively. So the total contributions of
these four figures are infrared finite. In other words, these vertex
corrections can be calculated without considering the transverse
momentum effects of the quark at the end-point region in collinear
factorization theorem. Therefore, there is no need to employ the
$k_T$ factorization theorem here. The vertex corrections to the
$B\to J/\psi K$ decays, denoted as $f_I$ in QCDF, have been
calculated in the NDR scheme \cite{ck00,cheng01}, and can be adopted
directly. Their effects can be combined into the Wilson coefficients
associated with the factorizable contributions: \beq
a_2&=&C_1+\frac{C_2}{N_c}+\frac{\alpha_s}{4\pi}\frac{C_F}{N_c}C_2
\left(-18+12\ln\frac{m_b}{\mu}+f_I\right)\;, \label{eq:a2}\eeq \beq
a_3&=&C_3+\frac{C_4}{N_c}+\frac{\alpha_s}{4\pi}\frac{C_F}{N_c}C_4
\left(-18+12\ln\frac{m_b}{\mu}+f_I\right)\;,\non
a_5&=&C_5+\frac{C_6}{N_c}+\frac{\alpha_s}{4\pi}\frac{C_F}{N_c}C_6
\left(6-12\ln\frac{m_b}{\mu}-f_I\right)\;, \label{eq:a35} \\
a_7&=&C_7+\frac{C_8}{N_c}+\frac{\alpha_s}{4\pi}\frac{C_F}{N_c}C_8
\left(6-12\ln\frac{m_b}{\mu}-f_I\right)\;,\non
a_9&=&C_9+\frac{C_{10}}{N_c}+\frac{\alpha_s}{4\pi}\frac{C_F}{N_c}C_{10}
\left(-18+12\ln\frac{m_b}{\mu}+f_I\right)\; ,\label{eq:a79}
\eeq
with the function $f_I$,
\begin{eqnarray}
f_I=\frac{2\sqrt{2N_c}}{f_{J/\Psi}}\int
dx_2\phi_{\jpsi}^L(x_2)\left[\frac{3(1-2x_2)}{1-x_2}\ln x_2-3\pi
i+3\ln(1-r_2^2)+\frac{2r_2^2(1-x_2)}{1-r_2^2 x_2}\right]\;,
\end{eqnarray}
where $r_2=m_{\jpsi}/m_B$ and those terms proportional to
$r_2^4$ have been neglected.
In Eqs.(\ref{eq:a2}-\ref{eq:a79}), the Wilson coefficients $C_i$ at NLO level
should be used when the NLO vertex corrections are taken into account.

For $B \to \eta_c K$ decays, it is easy to obtain the corresponding
NLO vertex corrections from those Wilson coefficients in
Eqs.~(\ref{eq:a2}-\ref{eq:a79}), by the replacement of
the parameter $f_{\jpsi}$ and $r_2(\jpsi)$ in $f_I$ with $f_{\eta_c}$ and $r_2(\eta_c)$
\cite{chao2}, respectively.

\begin{figure}[tb]
\begin{center}
\includegraphics[scale=0.65]{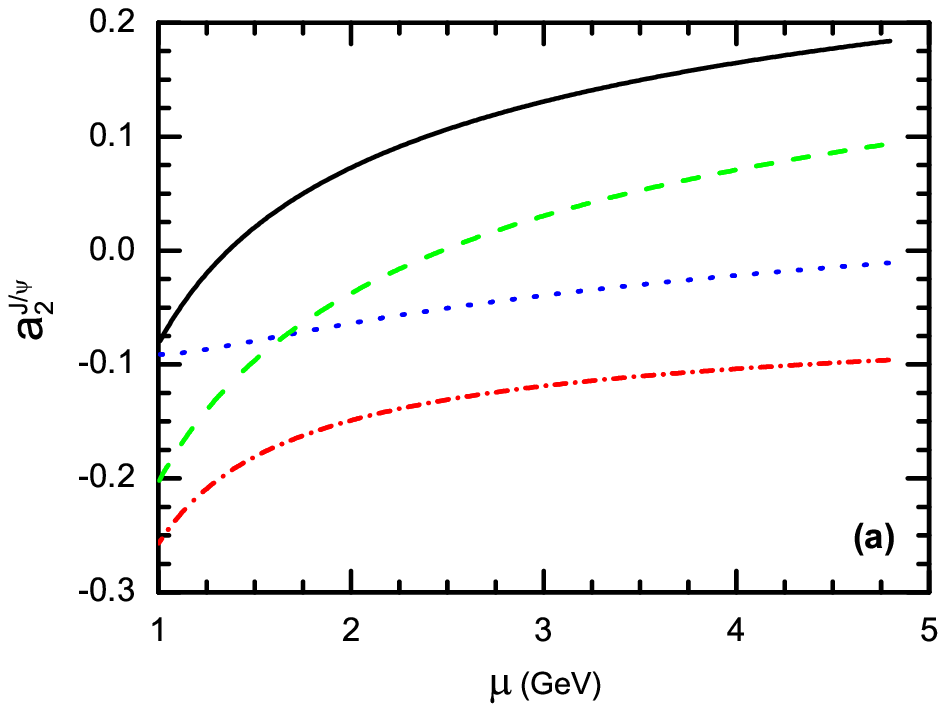}
\includegraphics[scale=0.65]{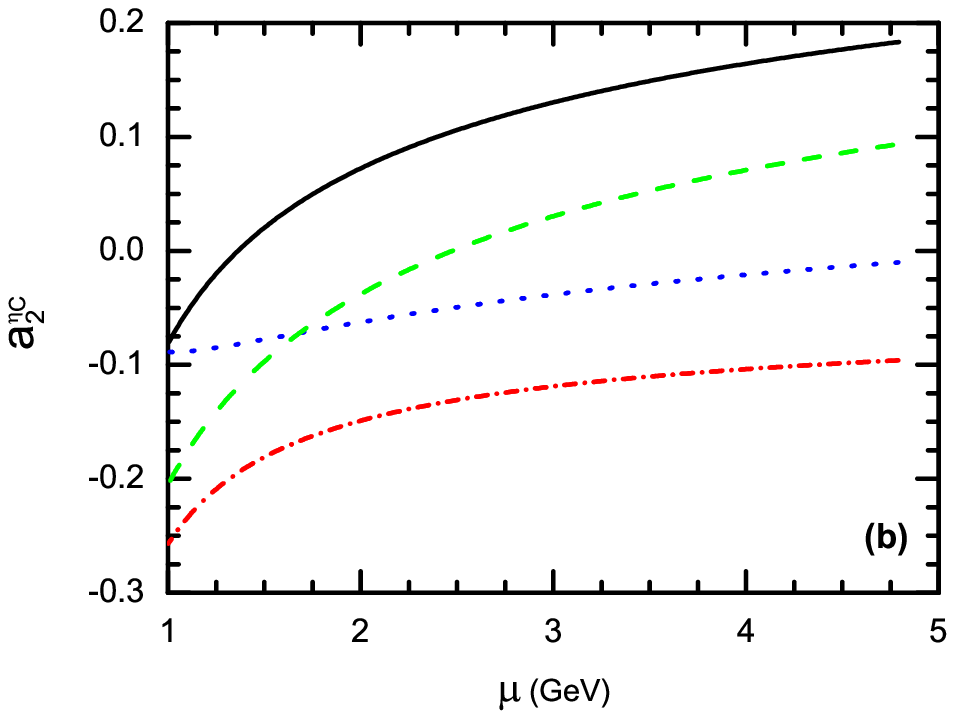}
\vspace{-0.3cm} \caption{Dependence of $a_{2}$ on the
renormalization scale $\mu$ for (a) $B \to \jpsi K$ and (b) $B \to \eta_c
K$ decays. The solid ( dashed ) curve stands for $a_{2}$ at NLO (LO) level without the
vertex corrections, while the dotted ( dash-dotted ) curve refers to the real
(imaginary) part of the $a_2$ at NLO level with  the vertex corrections.}
\label{fig:fig3}
\end{center}
\end{figure}

Since both $B \to \jpsi K$ and $\eta_c K$ are color-suppressed decays, the Wilson coefficient
$a_2$ in general plays the dominate role. It is instructive to check the variation of
$a_2$ at LO or NLO level, with or without the inclusion of the NLO vertex corrections.
From Figs.3a and 3b, one can see that (a) the $\mu$-dependence of
$a_2(\jpsi K)$ and $a_2(\eta_c K)$ are very similar;  (b)
the $\mu$-dependence of $a_2$ is decreased effectively due to the
inclusion of the NLO vertex corrections; and (c) the vertex correction provides a large
imaginary part to $a_2$, and therefore an effective enhancement of
$|a_2|$ due to the vertex correction is expected.

\section{Numerical results and Discussions}\label{sec:s3}

\subsection{Branching ratios}

The input parameters and the wave functions to be used in the
numerical calculations are given in Appendix \ref{sec:app2}.

Firstly, we find the pQCD predictions for the corresponding form
factors at zero momentum transfer:
\beq
F^{B\to K}_0(q^2=0)&=& 0.32^{+0.05}_{-0.05} (\omega_b)\;,
\eeq
for $f_B = 0.19$ GeV, and  $\omega_b=0.40\pm 0.04$GeV.
It agrees very well with that obtained in QCD sum rule
calculations\cite{melic1}.

Now we calculate the branching ratios for those considered decay
modes. With the complete decay amplitudes, we can obtain the decay
width for the considered decays,
\beq
\Gamma(B \to M_2 K) =\frac{G_F^2 m_B^3}{32\pi}(1-r^2)
\left| {\cal M} (B \to M_2 K)\right|^2,
\eeq
where $r=m_{\jpsi}/m_B$ or $m_{\eta_c}/m_B$.

By using the input parameters and wave functions as given in
Appendix \ref{sec:app2}, we find the LO and NLO pQCD predictions (in
unit of $10^{-4}$) for the CP-averaged branching ratios of the four
$B \to \jpsi K$ and $\eta_c K$ decays and show them in the
Table~\ref{tab:br}. The predictions listed in the column one are the
LO pQCD predictions by setting $\mu_0=1.0$ GeV. The predictions
listed in the column two is the NLO pQCD prediction by setting
$\mu_0=1.0$. The first theoretical error in these entries arises
from the B meson wave function shape parameter $\omega_b=0.40 \pm
0.04$ and the decay constants $f_{\jpsi}= 0.405 \pm 0.014$ GeV
and/or $f_{\eta_c}= 0.420 \pm 0.050$ GeV. The second error is from
the combination of the uncertainties of Gegenbauer moments
$a_1^K=0.17\pm 0.17$ and/or $a_2^{K}=0.115\pm 0.115$. In the fourth
column of Table~\ref{tab:br}, as a comparison, we also cite the
typical theoretical predictions obtained previously by using various
approaches or models \cite{ck00,cheng01,chao2,melic}.

In the last column of Table~\ref{tab:br}, we list the world averages
of the experimental measurements \cite{pdg2008,hfag2008}. One can
see that the LO pQCD predictions for the branching ratios are indeed
much smaller than the measured values, but the pQCD predictions with
the inclusion of the vertex corrections can enhance the branching
ratios evidently for both $B \to \jpsi K$ and $B \to \eta_c K$
decays. When the NLO enhancement are included, the pQCD predictions
are basically consistent with the data within the still large theoretical errors.
Of course,  the central values are still about 40\% smaller than the measured values.
There is still some room left for non-perturbative contributions.

\begin{table}[htb]
\caption{The LO and NLO pQCD predictions (in unit of $10^{-4}$ )
of $Br(B \to \jpsi K)$ and $Br(B \to \eta_c K)$. For comparison, we also cite
the typical theoretical predictions as given in previous literatures
(in the fourth column), and the measured values
\cite{pdg2008,hfag2008}. }
\label{tab:br}
\begin{center}\vspace{-0.5cm}
\begin{tabular}[t]{|l|c|c|l|l|} \hline  \hline
Channels &   LO  &NLO            & Others &  Data \\ \hline
 $B^0 \to J/\Psi K^0 $ &$1.1^{+0.8+0.9}_{-0.5-0.5}$ &$5.2^{+1.0+3.4}_{-0.9-2.6}$& $\sim 1.0$ \cite{ck00} & $8.71  \pm 0.32$ \\
 $B^+ \to J/\Psi K^+ $ &$1.2^{+0.9+0.9}_{-0.5-0.5}$ &$5.6^{+1.0+3.6}_{-0.9-2.8}$& $\sim 3.3$ \cite{melic} & $10.07\pm 0.35$ \\ \hline
 $B^0 \to \eta_c K^0 $ &$0.8^{+0.5+0.1}_{-0.3-0.1}$ &$5.5^{+2.1+1.0}_{-1.7-1.0}$ & $\sim 2 $\cite{chao2,chen05} & $8.9 \pm 1.6$ \\
 $B^+ \to \eta_c K^+ $ &$0.8^{+0.5+0.2}_{-0.2-0.1}$ &$5.9^{+2.2+1.2}_{-1.8-1.1}$ & $\sim 2$\cite{melic,chen05}  & $9.1 \pm 1.3 $ \\ \hline
\hline
\end{tabular}
\end{center}
\end{table}

Now we investigate in more detail why the vertex corrections can provide a
significant enhancement.
The total decay amplitudes as given in Eqs.(\ref{eq:jpsim3}) and (\ref{eq:etacm3})
can be in re-written in the following form
\beq
M&=&\left [ F_C + F_P\right]_{Fac.} + \left [ M_T + M_P\right]_{Spec.},
\label{eq:mfs}
\eeq
where $F_C$ and $F_P$ stands for the ``color-suppressed" and the penguin part
of the factorizable contribution, coming from the emission diagram Fig.1a and 1b;
while $M_T$ and $M_P$ stands for the ``Tree"  and the penguin part
of the nonfactorizable contribution, coming from the spectator diagram Fig.1c and 1d.
From Eq.~(\ref{eq:jpsim3}), for example, it ie easy to separate the total decay amplitude
$\calm (B \to \jpsi K)$ into the following four parts
\beq
F_C&=&F_{J/\Psi K}^{V-A} f_{J/\Psi}\; V_{cb}^*V_{cs}\; a_2; \non
F_P&=&- F_{J/\Psi K}^{V-A} f_{J/\Psi} \; V_{tb}^{*}V_{ts}\;  \left ( a_3+a_5+a_7 +a_9 \right);
\label{eq:fp01}\\
M_T&=& M_{J/\Psi K}^{V-A} \; V_{cb}^*V_{cs}\;  C_2; \non
M_P&=& - M_{J/\Psi K}^{V-A} \; V_{tb}^*V_{ts}\;   \left (C_4-C_6-C_8+C_{10}\right)
\label{eq:mp01}
\eeq

By numerical calculations, we find easily the
numerical values (in unit of $10^{-3}$) of the individual parts and
the total decay amplitude of $\calm(B \to \jpsi K)$ at the LO and
NLO level:
\beq
\calm^{LO} &=& \underbrace{-1.147}_{F_C} -
\underbrace{0.092}_{F_P} + \underbrace{(1.487 + i 0.529)}_{M_T} +
\underbrace{(0.046+ i 0.004)}_{M_P},\non
&=& 0.294 + i 0.533, \label{eq:mlo01}\\
\calm^{NLO} &=& \underbrace{(-0.546 -i 1.433)}_{F_C} +
\underbrace{(0.004 -i 0.038)}_{F_P} + \underbrace{(1.367 + i
0.485)}_{M_T} + \underbrace{(0.037+ i 0.002)}_{M_P},\non
&=& 0.862 - i 0.984, \label{eq:mnlo01}
\eeq
and the ratio of the square of the
decay amplitude $\calm^{NLO}$ and $\calm^{LO}$ is
\beq
R_M(B\to \jpsi K )=\frac{\left| \calm^{NLO}\right|^2}{\left|
\calm^{LO}\right|^2 } =4.62. \label{eq:rm288}
\eeq
For $B \to \eta_c K $ decay, we find the similar result: $R_M(B\to \eta_c K)=7.0$.

From the above numerical results, it is easy to see that
\begin{itemize}
\item
As generally expected, both the the factorizable penguin contribution ( $F_P$)
and the nonfactorizable part ( $M_P$) to the total decay amplitude are always
small in magnitude ( less than $10\%$), when compared with the "Tree" and "Color-suppressed"
parts ($M_T$ and $F_C$).

\item
At the leading order, $F_C=-1.147$ is large in size, but largely
canceled by the real part of $M_T$ ( ${\rm Re}(M_T)=1.487$ )
when one sums up the factorizable and nonfactorizable
contributions.
This strong cancelation results in a small LO pQCD prediction for the decay rates.

\item
At the next-to-leading order, the NLO Wilson coefficients will be used.
The penguin part $F_P$ and $M_P$ remain small in magnitude. The variations of
$M_T$ and $M_P$ due to the replacement of the LO Wilson coefficients by the NLO ones
are also small as expected\footnote{In the evaluation of $\calm^{LO}$ and $\calm^{NLO}$,
the LO and NLO Wilson coefficients $C_2$ and $C_{4,6,8,10}$ will be used, respectively.}.
For the ``color-suppressed" $F_C$ part, however, things become much different.
The real part of $F_C$ changes from $-1.147$ to $-0.546$, the previous large cancelation
between the real parts of $F_C$ and $M_T$ become weak significantly, while
a large imaginary part ${\rm Im}(F_C)=-1.433$ is also produced. These two changes
lead to a large $|\calm^{NLO}|^2$ and consequently a large NLO pQCD prediction
of the branching ratios.

\item
Although the NLO pQCD predictions for the branching ratios of the considered decays 
are consistent with the date within the still large theoretical uncertainties, 
but the central values of the NLO pQCD predictions, as listed in Table \ref{tab:br},  
are still about $60\%$ of the measured values. 
Certainly, there are still some room left for the non-perturbative long distance effects or 
other unknown  high order corrections.

\item
Among the three kinds of known NLO contributions in the pQCD approach, 
only the vertex corrections are relevant to  $B \to (\jpsi,\eta_c) K $ decays and taken into account 
here.
Other possible NLO contributions coming from the Feynman diagrams
as shown in Figs.5-7 in Ref.~\cite{xiao08} are still unknown at present. But they are 
generally expected to be the small part of the NLO contributions in the 
pQCD factorization approach\cite{nlo05,xiao08}.

\item
In Ref.~\cite{plb542}, the authors attempted to estimate  the soft
rescattering effect for $B^+ \to \jpsi K^+$ decay and concluded
that such effect is comparable in size with the experimental one. 
But it is worth mentioning that the estimation as presented in Ref.~\cite{plb542} 
has a very large theoretical uncertainty. 
We believe that the long distance effects for the considered decays are exist and may be large, 
but  much more careful studies should be made before one can find an reliable estimation 
about them \cite{cheng01,chen05,li07a}.

\end{itemize}


As discussed in previous section, we choose $\mu_0=1.0$ GeV as the
lower cut-off of the hard scale $t$. It should be noted that in the
considered decay channels, the characteristic hard energy scale
maybe smaller than 1 GeV, while all the meson distribution
amplitudes are defined at 1 GeV. In the numerical integration, we
therefore should set the Wilson coefficients $C_i(t)=C_i(1\;$GeV$)$
whenever $t \leq 1$ GeV to ensure the reliability of our
perturbative calculations.


Now we turn to study the ratios of the branching ratios for some phenomenologically
relevant decay modes. One advantage of the ratios is the possible cancelation
of the theoretical uncertainties of individual calculations.
Using the data of the measured branching ratios as given in Ref.~\cite{pdg2008},
the three ratios $R_i^{exp}$ can be defined as the following
\beq
R_1^{exp}&=& \frac{Br(B^+ \to \eta_c K^+)}{Br(B^+ \to \jpsi K^+)}  =0.90 \pm 0.13,\non
R_2^{exp}&=& \frac{Br(B^0 \to \eta_c K^0)}{Br(B^0 \to J/\Psi K^0)} = 1.02 \pm 0.19,\non
R_3^{exp}&=& \frac{Br(B^0 \to \eta_c K^0)}{Br(B^- \to \eta_c K^-)} = 0.88 \pm 0.16 \;.
\label{eq:r123}
\eeq

\begin{table}[htb]
\caption{The ratios $R_i$ of the pQCD predictions for the branching
ratios. The ratios as given in PDG 2008 \cite{pdg2008} are also 
shown in the last column.} 
\label{tab:ratios}
\begin{center}\vspace{-0.5cm}
\begin{tabular}[t]{|c|c|c|c|} \hline  \hline
Ratios   & LO    & NLO &  Data \\
\hline
$R_1 $  & $0.7^{+0.9}_{-0.4} $    &$1.05^{+1.14}_{-0.42}$  &$0.90  \pm 0.13$ \\
$R_2 $  & $0.7^{+1.3}_{-0.3} $    &$1.06^{+1.23}_{-0.43}$  &$1.02  \pm 0.19$ \\
$R_3 $  & $1.0^{+0.6}_{-0.4} $    &$0.93^{+0.52}_{-0.27}$  &$0.88  \pm 0.16$ \\
\hline \hline
\end{tabular}
\end{center}
\end{table}

By comparing the pQCD predictions of the ratios and the measured
ones, as listed in Table \ref{tab:ratios}, 
we can see that (a) the consistency between the pQCD prediction for the ratios and the 
measured ones is improved significantly when the NLO contributions are taken into account;
(b) the theoretical uncertainties are still large because of our poor knowledge about the
Gegenbauer moments $a_{1,2}^K$. 

In short, from the above pQCD predictions for the branching ratios
and the detailed phenomenological analysis, we can conclude that
the pQCD predictions for the branching ratios become close to the data due to the 
significant enhancement of the NLO vertex corrections.

\subsection{CP-violating asymmetries}

Now we turn to the evaluations of the CP-violating asymmetries of $B
\to M_2\;M_3$ decays in pQCD approach. For the charged B meson
decays,the direct CP-violating asymmetries $\acp^{dir}$ can be
defined as usual. For both $B^+ \to \jpsi K^+$ and $\eta_c K^+$
decays, there are no direct CP violation, since there is no weak
phase appeared in their decay amplitude, as can be seen easily in
Eqs.~(\ref{eq:jpsim3}) and (\ref{eq:etacm3}). This theoretical expectation 
agrees well with the data \cite{pdg2008,hfag2008}: 
\beq
\acp^{dir}(B^+ \to J/\Psi K^+) &=& 0.017 \pm 0.016 \;,\non
\acp^{dir}(B^+\to \eta_c K^+) &=&  -0.16 \pm 0.08\;.
\eeq
which are consistent with zero within $2\sigma$ errors.

For the $B^0 \to M_2 M_3$ decays, because these decays are
neutral B meson decays, so we should consider the effects of
$B^0-\bar{B}^0$ mixing. The direct and mixing induced CP-violating asymmetries
$\acp^{dir}$ and $\acp^{mix}$ can be written as
\beq
\acp^{dir}= \frac{ \left | \lambda_{CP}\right |^2 -1 } {1+|\lambda_{CP}|^2}, \quad
\acp^{mix}= \frac{ 2Im (\lambda_{CP})}{1+|\lambda_{CP}|^2},
\label{eq:acp-dm}
\eeq
where the CP-violating parameter $\lambda_{CP}$ is
\beq
\lambda_{CP} = \eta_f \frac{ V_{tb}^*V_{td} \langle f |H_{eff}|
\overline{B}^0\rangle} { V_{tb}V_{td}^* \langle f |H_{eff}|B^0\rangle}
= \eta_f e^{-2 i \beta} \frac{\langle f |H_{eff}| \overline{B}^0\rangle}{\langle f |H_{eff}|
B^0\rangle} , \label{eq:lambda2}
\eeq
where $\eta_f$ is the CP-eigenvalue of the final states.
By using the the input parameters as given in Appendix B, we find  the 
following pQCD predictions
\beq
\acp^{dir}(B^0 \to \jpsi K_S^0)&=& \acp^{dir}(B^0 \to \eta_c K_S^0)\approx 0 \;, \non
\acp^{mix}(B^0 \to \jpsi K_S^0)&=& \acp^{mix}(B^0 \to \eta_c K_S^0)=
\left (70.9 ^{+2.8} _{-2.7} \right )\%,
\label{eq:cpv}
\eeq
where the dominant error comes from $\bar{\rho}=0.135^{+0.031}_{-0.016}$ and
$\bar{\eta}=0.349^{+0.015}_{-0.017}$ \cite{pdg2008}. It is easy to see that
the pQCD predictions for CP -violating asymmetries agree perfectly with the experimental
measurements \cite{pdg2008,hfag2008}.

\section{Summary}

In this paper, we calculated the branching ratios and CP-violating
asymmetries of the four $B \to (\jpsi,\eta_c) K$ decays
by employing the pQCD factorization approach with the inclusion of currently
known NLO contributions.

From our numerical calculations and phenomenological analysis, we found the
following results:
\begin{itemize}
\item
The inclusion of the known NLO  contributions can result in a factor of five enhancements to the leading order 
results. The NLO pQCD predictions for the branching ratios are the following
\beq
Br(B^0 \to \jpsi K^0) &=& 5.2^{+3.5}_{-2.8}\times 10^{-4}, \non
Br(B^+ \to \jpsi K^+) &=& 5.6^{+3.7}_{-2.9}\times 10^{-4}, \non
Br(B^0 \to \eta_c K^0) &=& 5.5^{+2.3}_{-2.0}\times 10^{-4}, \non
Br(B^+ \to \eta_c K^+) &=& 5.9^{+2.5}_{-2.1}\times 10^{-4}. 
\eeq
Although the central values of the pQCD predictions are still $40\%$ smaller than the measured ones, 
they basically agree with the data within $2\sigma$ errors. One can also see that, on the other hand, 
there are still some room left for the non-perturbative contributions.

\item
The pQCD predictions for the CP-violating asymmetries of the considered decays
also agree perfectly with the data.

\item
In this paper, only those currently known NLO contributions
have been taken into account. To obtain a complete NLO calculations in the pQCD approach, 
the still missing pieces should be evaluated as soon as possible.

\end{itemize}

\begin{acknowledgments}

The authors are very grateful to Hsiang-nan Li, Cai-Dian L\"u and Ying Li
for valuable discussions. This work is partially supported by the National
Natural Science Foundation of China under Grant No.10575052, 10605012 and
10735080.

\end{acknowledgments}


\begin{appendix}

\section{Related Functions }\label{sec:app1}

We show here the hard function $h_i$'s, coming from the Fourier
transformations  of the function $H^{(0)}$,
\beq
h_e(x_1,x_3,b_1,b_3)&=&
K_{0}\left(\sqrt{x_1 x_3(1-r^2)} m_{B} b_1\right)
\left[\theta(b_1-b_3)K_0\left(\sqrt{x_3(1-r^2)} m_{B}
b_1\right)\right.
 \non
& &\;\left. \cdot I_0\left(\sqrt{x_3(1-r^2)} m_{B}
b_3\right)+\theta(b_3-b_1)K_0\left(\sqrt{x_3(1-r^2)}  m_{B}
b_3\right)\right.
 \non
& &\;\left. \cdot I_0\left(\sqrt{x_3(1-r^2)}  m_{B}
b_1\right)\right] S_t(x_3), \label{he1} \eeq 
 \beq
 h_{f}(x_1,x_2,x_3,b_1,b_2) &=&
 \biggl\{\theta(b_2-b_1) \mathrm{I}_0(m_{B}\sqrt{x_1 x_3(1-r^2)} b_1)
 \mathrm{K}_0(m_{B}\sqrt{x_1 x_3(1-r^2)} b_2)
 \non
&+ & (b_1 \leftrightarrow b_2) \biggr\}  \cdot\left(
\begin{matrix}
 \mathrm{K}_0(m_{B} F_{(1)} b_2), & \text{for}\quad F^2_{(1)}>0 \\
 \frac{\pi i}{2} \mathrm{H}_0^{(1)}(m_{B}\sqrt{|F^2_{(1)}|}\ b_2), &
 \text{for}\quad F^2_{(1)}<0
\end{matrix}\right),
\label{eq:pp1}
 \eeq
where $J_0$ is the Bessel function, $K_0$ and $I_0$ are the modified
Bessel functions with  $K_0 (-i x) = -(\pi/2) Y_0 (x) + i (\pi/2)
J_0 (x)$, and $F_{(1)}^2$ is defined by \beq F^2_{(1)}&=&(x_1 -x_2)
(x_3+(x_2-x_3)r^2)+r_c^2\;.
 \eeq
The threshold resummation form factor $S_t(x_i)$ can be found in
Ref.~\cite{tls}.

The Sudakov factors used in the text are defined as
\beq
S_{ab}(t)
&=& s\left(x_1 m_{B}/\sqrt{2}, b_1\right) +s\left(x_3
m_{B}/\sqrt{2}, b_3\right) +s\left((1-x_3) m_{B}/\sqrt{2},
b_3\right) \non
&&-\frac{1}{\beta_1}\left[\ln\frac{\ln(t/\Lambda)}{-\ln(b_1\Lambda)}
+\ln\frac{\ln(t/\Lambda)}{-\ln(b_3\Lambda)}\right],
\label{wp}\\
S_{cd}(t) &=& s\left(x_1 m_{B}/\sqrt{2}, b_1\right)
 +s\left(x_2 m_{B}/\sqrt{2}, b_2\right)
+s\left((1-x_2) m_{B}/\sqrt{2}, b_2\right) \non
 && +s\left(x_3
m_{B}/\sqrt{2}, b_1\right) +s\left((1-x_3) m_{B}/\sqrt{2},
b_1\right) \non
 & &-\frac{1}{\beta_1}\left[2
\ln\frac{\ln(t/\Lambda)}{-\ln(b_1\Lambda)}
+\ln\frac{\ln(t/\Lambda)}{-\ln(b_2\Lambda)}\right], \label{Sc} \eeq
where the function $s(q,b)$ are defined in the Appendix A of
Ref.~\cite{luy01}. The scale $t_i$'s in the above equations are
chosen as \beq t_{e}^1 &=&  {\rm
max}(\sqrt{x_3(1-r^2)}m_{B},1/b_1,1/b_3), \non t_{e}^2 &=& {\rm
max}(\sqrt{x_1(1-r^2)}m_{B},1/b_1,1/b_3),\non t_{f} &=& {\rm
max}(\sqrt{x_1 x_3(1-r^2)}m_{B},\sqrt{(x_1-x_2)x_3(1-r^2)+r_c^2}m_{B},
1/b_1,1/b_2),
\eeq
where $r=m_{M_2}/m_{B} (M_2= J/\Psi,\eta_c)$,$r_c=m_c/m_B$.
The scale $t_i$'s are chosen as the
maximum energy scale appearing in each diagram in order to kill the large
logarithmic radiative corrections.

\section{Input parameters and wave functions} \label{sec:app2}

The masses, decay constants, QCD scale  and $B$ meson lifetime are
the following
 \beq
f_{J/\Psi} &=& 0.405 {\rm GeV},\quad f_K = 0.16 {\rm GeV} , \quad
f_{\eta_c} = 0.42 {\rm GeV}, \quad m_W = 80.41{\rm GeV}, \non
m_{\eta_c} &=& 2.98  {\rm GeV}, \quad m_{B} = 5.2794  {\rm GeV}, \quad
 m_{J/\Psi} = 3.097 {\rm GeV}, \non
\tau_{B^+}&=& 1.643 {\rm  ps},\quad \tau_{B^0}=1.53 {\rm  ps}.
\label{para}
\eeq

For the CKM matrix elements, here we adopt the Wolfenstein
parametrization for the CKM matrix, and take $\lambda=0.2257,
A=0.814, \bar{\rho}=0.135$ and $\bar{\eta}=0.349$
\cite{pdg2008}.

As for $B$ meson wave function, we make use of the same
parameterizations as in Ref.~\cite{luy01}.
We adopt the model
\beq
\phi_{B}(x,b) &=& N_{B}
x^2(1-x)^2 \mathrm{exp} \left
 [ -\frac{m_{B}^2\ x^2}{2 \omega_{b}^2} -\frac{1}{2} (\omega_{b} b)^2\right],
 \label{phib}
\eeq
where $\omega_{b}$ is a free parameter and we take $\omega_{b}=0.40\pm 0.04$ GeV in numerical
calculations, and $N_{B}=91.745$ is the normalization factor for $\omega_{b}=0.40$.

For the vector $J/\Psi$ meson, we take the wave function as follows,
\beq \Phi_{J/\Psi}(x) &=&\frac{1}{\sqrt{2N_c}}
\bigg\{m_{J/\Psi}\epsl_L \phi_{J/\Psi}^L(x)+\epsl_L \psl
\phi_{J/\Psi}^{t}(x) \bigg\}\;. \eeq Here, $\phi^L$ denote for the
twist-2 DA's, and $\phi^t$ for the twist-3 ones, both of them have
experimental and theoretical basis\cite{bc04}. $x$ represents the
momentum fraction of the charm quark inside the charmonium. The
$J/\Psi$ meson asymptotic distribution amplitudes read as
\cite{bc04} \beq
\phi_{J/\Psi}^L(x)&=&9.58\frac{f_{J/\Psi}}{2\sqrt{2N_c}}x(1-x)
\left[\frac{x(1-x)}{1-2.8x(1-x)}\right]^{0.7}\;,\non
\phi_{J/\Psi}^t(x)&=&10.94\frac{f_{J/\Psi}}{2\sqrt{2N_c}}(1-2x)^2
\left[\frac{x(1-x)}{1-2.8x(1-x)}\right]^{0.7}\;.\label{jda} \eeq It
is easy to see that both the twist-2 and twist-3 DA's vanish at the
end points due to the factor $[x(1-x)]^{0.7}$.

For pseudoscalar meson $\eta_c$, the wave function is  the
form of
\beq
\Phi_{\eta_c}(x)&=& \frac{i}{\sqrt{2 N_c}}\gamma_5
\bigg\{\psl \phi_{\eta_c}^v+m_{\eta_c} \phi_{\eta_c}^s\bigg\}\;.
\eeq
The twist-2 and twist-3 asymptotic distribution
amplitudes, $\phi^v$ and $\phi^s$, can be written
as\cite{bc04},
\beq
\phi_{\eta_c}^v(x)&=&9.58\frac{f_{\eta_{c}}}{2\sqrt{2N_c}}x(1-x)
\left[\frac{x(1-x)}{1-2.8x(1-x)}\right]^{0.7}\;, \non
\phi_{\eta_c}^s(x)&=&1.97\frac{f_{\eta_{c}}}{2\sqrt{2N_c}}
\left[\frac{x(1-x)}{1-2.8x(1-x)}\right]^{0.7}\;.
 \label{eda}
 \eeq

The twist-2 kaon distribution amplitude $\phi^A_{K}$, and the
twist-3 ones $\phi_{K}^P$ and $\phi_{K}^T$ have been parameterized
as \cite{pball06}
\begin{eqnarray}
\phi_{K}^A(x) &=& \frac{f_{K}}{2\sqrt{2N_c}}\, 6x(1-x) \left[1 +
a_1^{K} C_1^{3/2}(t) + a_2^{K}C_2^{3/2}(t) \right. \non & &
 \hspace{35mm} \left. +a_4^{K}C_4^{3/2}(t)\right] ,
\\
\phi^P_{K}(x) &=& \frac{f_{K}}{2\sqrt{2N_c}}\, \bigg[ 1
+\left(30\eta_3 -\frac{5}{2}\rho_{K}^2\right) C_2^{1/2}(t)
\nonumber\\
& & \hspace{25mm} -\, 3\left\{ \eta_3\omega_3 +
\frac{9}{20}\rho_{K}^2(1+6a_2^{K}) \right\} C_4^{1/2}(t)\bigg]\;,
\\
\phi^T_{K}(x) &=& \frac{f_{K}}{2\sqrt{2N_c}}\, (1-2x)\bigg[ 1 +
6\left(5\eta_3 -\frac{1}{2}\eta_3\omega_3 - \frac{7}{20}
      \rho_{K}^2 - \frac{3}{5}\rho_{K}^2 a_2^{K} \right)
 \non & &
 \hspace{35mm} \times (1-10x+10x^2) \bigg]\;,\ \ \ \
\end{eqnarray}
with $t=2x -1$, the Gegenbauer moments $a_1^K=0.17$,
$a^{K}_2=0.115$, and $a_4^K=-0.015$, the parameters $\eta_3=0.015$,
$\omega_3=-3.0$, the mass ratio
$\rho_{K}=(m_{u(d)}+m_{s})/m_{K}=m_{K}/m_{0K}$ and the Gegenbauer
polynomials $C_n^{\nu}(t)$, \beq C_1^{3/2}(t)&=& 3t,\quad
C_2^{1/2}(t)= \frac{1}{2} \left(3 t^2-1\right),\quad C_2^{3/2}(t)=
\frac{3}{2} \left(5t^2-1\right), \non C_4^{1/2}(t)&=& \frac{1}{8}
\left(3-30 t^2+35t^4\right), \quad C_4^{3/2}(t)=\frac{15}{8}
\left(1-14 t^2+21 t^4\right). \eeq

\end{appendix}


\end{document}